\documentclass[%
 reprint,
 amsmath,amssymb,
 aps,
]{revtex4-1}
\usepackage{graphicx}% Include figure files
\usepackage{dcolumn}% Align table columns on decimal point
\usepackage{bm}% bold math
\begin{document}

\preprint{APS/123-QED}

\title{Phonon dispersion and electron-phonon interaction
in peanut-shaped fullerene polymers
}% Force line breaks with \\

\author{Shota Ono$^*$ and Hiroyuki Shima}

\affiliation{%
Division of Applied Physics, Faculty of Engineering,
Hokkaido University, Sapporo, Hokkaido 060-8628, Japan
}%

\date{\today}% It is always \today, today,

\begin{abstract}
We reveal that the periodic radius modulation peculiar to one-dimensional (1D) peanut-shaped fullerene (C$_{60}$) polymers exerts a strong influence on their low-frequency phonon states and their interactions with mobile electrons. The continuum approximation is employed to show the zone-folding of phonon dispersion curves, which leads to fast relaxation of a radial breathing mode in the 1D C$_{60}$ polymers. We also formulate the electron-phonon interaction along the deformation potential theory, demonstrating that only a few set of electron and phonon modes yields a significant magnitude of the interaction relevant to the low-temperature physics of the system. The latter finding gives an important implication for the possible Peierls instability of the C$_{60}$ polymers suggested in the earlier experiment.

\begin{description}
\item[PACS numbers]
72.80.Rj, 63.22.-m, 63.20.kd
%use the optional argument of the \verb+\item+ command to give the category of e%ach item. 
\end{description}
\end{abstract}

\pacs{Valid PACS appear here}% PACS, the Physics and Astronomy
\maketitle

%\tableofcontents

\section{\label{sec:level1} INTRODUCTION}
Successful syntheses of fullerene (C$_{60}$) molecules 
and related nanomaterials \cite{rao1,rao2} have inspired tremendous interest
in the community of nanotechnology and nanoscience.
Experimental studies have shown that electron-beam
irradiation to a C$_{60}$ film induces polymerizations and produces
one-dimensional (1D) C$_{60}$ polymers \cite{onoe1,onoe2}.
The polymers have peanut-like geometry, i.e.,
the tube structure with periodically modulated radius along the
axis \cite{wang,beu1,beu2,hnaka,takashima1}.
Such the peanut structure was confirmed by comparison between
the infrared spectral measurement \cite{onoe1} and the first-principle
density-functional calculations of the vibrational
spectra \cite{beu2,takashima1}.
Recently, it was predicted that the periodic radius modulation strongly affects
transport properties of the carrier in the 1D C$_{60}$ polymers \cite{shima1}.
Similar periodic corrugation of a two-dimensional conducting film was also
shown to increase the electrical resistivity \cite{ono1}.
These findings are attributed to an effective electrostatic field caused by
the periodic corrugation, which gives rise to a Fermi velocity shift
and electron-electron scattering enhancement.

The 1D C$_{60}$ polymers exhibit semiconducting or metallic nature depending on the irradiation time interval \cite{onoe2}. This suggests that well-controlled irradiation time may lead the fabrication of quantum wires with desired conductivity. Of particular interest is the metallic C$_{60}$ polymers; in the metallic cases, nonequiliblium photo-excited carriers exhibit anomalously slow relaxation below 60K, indicating pseudogap formation in the electron energy band near the Fermi level \cite{toda}. Phenomenologically, this pseudogap is likely to result from the Peierls instability associated with a periodic distortion of
the underlying carbon structure. However, its theoretical understanding has not been attained; in fact, it has remained unclear how the periodic radius modulation in the C$_{60}$ polymers affects the phononic excitations and electron-phonon couplings in the systems.

We point out that the periodic geometry of the 1D C$_{60}$ polymers will induce a pronounced shift in the low-frequency phonon spectrum where the phonon wavelength is comparable to the radius modulation period, i.e., larger enough than the carbon bond length. In the low-frequency range, continuum elastic approximation is a powerful means to analyze the phononic excitations. Furthermore, the approximation allows to use the deformation potential theory \cite{kittel} that describes the electron-phonon interaction in a conducting medium. Combining the two methods, therefore, we can formulate the radius modulation effect on the phonon spectrum and the electron-phonon interaction in the 1D C$_{60}$ polymers; the results obtained give a fundamental clue to explore novel properties of collective excitations in the system.

In this paper, we study the low-frequency phonons and their coupling with electrons in the 1D C$_{60}$ polymers by using a continuum model. A generalized eigenvalue equation is formulated to obtain phonon dispersion relations that show zone-folding due to the periodic geometry. It is demonstrated that only a particular set of phonon modes, among many choices, yields marked enhancement in the electron-phonon coupling strength; these specific phonon modes are thought to play a prominent role in the possible Peierls instability in the 1D C$_{60}$ polymers. We also show that the zone-folding in the phonon spectrum encourages the relaxation of radial breathing modes in the 1D C$_{60}$ polymers, which is a consequence of the periodic radius modulation.

In Sec.~\ref{subsec:level2_a}, we derive an eigenvalue equation that is useful for calculating eigenmodes of general isotropic elastic medium with cylindrical symmetry. In Sec.~\ref{subsec:level2_b}, we introduce a continuum elastic model of the 1D C$_{60}$ polymers and suitable basis functions for an efficient eigenmode analysis: Appropriate values of the elastic constants and geometric parameters of the model are presented in Sec.~\ref{subsec:level2_c}. Section~\ref{sec:level3} shows the zone-folding of phonon dispersion curves caused by the periodic geometry. Section.~\ref{sec:level4} is devoted to formulate the electron-phonon interaction Hamiltonian on the basis of the deformation potential theory. Numerical results of the coupling strength are shown in Sec.~\ref{sec:level5}, and associated physical implications are discussed in Sec.~\ref{sec:level6}. The paper is closed by conclusion given in Sec.~\ref{sec:level7}.
%%%%%%%%%%%%%%%%%%%%%%%%%%%%    GRAPHICS   %%%%%%%%%%%%%%%%%%%%%%%%%%%%%%
\begin{figure}[ttt]
\center
\includegraphics[scale=0.5,clip]{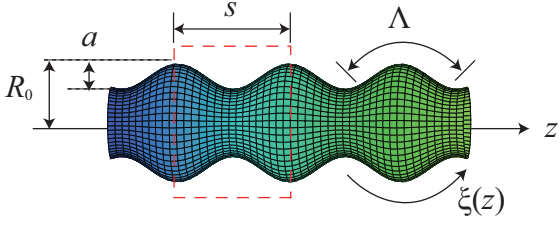}%-----------------------------%
\caption{\label{fig:tube}(Color online)
Continuum model of a 1D peanut-shaped C$_{60}$ polymer. The dashed square (red in color) indicates the unit cell.}
\end{figure}
%%%%%%%%%%%%%%%%%%%%%%%%%%%%%%%%%%%%%%%%%%%%%%%%%%%%%%%%%%%%%%%%%%

\section{\label{sec:level2} CONTINUUM ELASTIC MODEL}
In this section, we formulate a continuum elastic model of the 1D C$_{60}$ polymers. Thus far, several continuum models have been developed to study low-frequency phonons in nanostructures such as GaAs/AlAs based nanowires \cite{mizuno,nishiguchi1}, pristine C$_{60}$ molecules \cite{kahn}, and carbon nanotubes \cite{mahan,suzuura,chico,li}. Our model discussed below takes into account the effects of the periodic structure on the vibrational properties of the system, and the resultant vibrational eigenmodes will be applied for defining the electron-phonon interaction Hamiltonian in Sec.~\ref{sec:level4}.

\subsection{\label{subsec:level2_a} EIGENVALUE EQUATION}
The outline below is our derivation of a generalized eigenvalue equation [i.e., Eq.~(\ref{eq:general})] that allows to obtain the vibrational eigenmodes of an elastic medium. For general cylindrical objects, elastic vibration is governed by the Lagrangian:
\begin{equation}%------------------------------------------%
 \mathcal{L}=\int_{V} \left( \frac{\rho \omega^2}{2} \sum_i u_i^* u_i
 -\frac{\lambda}{2} \Delta^* \Delta-\mu \sum_i \sum_j
 e_{ij}^*e_{ij} \right)dV. \label{eq:lagrangian}
\end{equation}%--------------------------------------------%
Here, $\rho$ is the mass density, $\omega$ is the vibrational frequency,
$\lambda$ and $\mu$ are Lame coefficients.
The complex-valued function $u_i$ denotes the $i$th component of
the displacement vector $\bm{u}(\bm{r})$ at a point $\bm{r}$,
and $u_{i}^{*}$ is its complex conjugate.
Using the cylindrical coordinates, we can write
$\Delta \equiv e_{rr}+e_{\theta\theta}+e_{zz}$
with the strain tensors \cite{landau}:
\begin{eqnarray}%------------------------------------------%
 e_{rr}&=& \partial_r u_r, \ \ 
 e_{\theta \theta}=\frac{1}{r}\partial_\theta u_\theta
                  +\frac{u_r}{r}, \ \ 
 e_{zz}=\partial_z u_z,
 \nonumber\\
 e_{r\theta}&=&e_{\theta r}=\frac{1}{2}\left(
    \partial_r u_\theta -\frac{u_\theta}{r}+\frac{1}{r}\partial_\theta u_r
   \right),
 \nonumber\\
 e_{\theta z}&=&e_{z\theta}=\frac{1}{2}\left(
           \frac{1}{r}\partial_\theta u_z+\partial_z u_\theta
   \right),
 \nonumber\\
 e_{zr}&=&e_{rz}=\frac{1}{2}\left(
   \partial_z u_r + \partial_r u_z \right), \label{eq:strain}
\end{eqnarray}%--------------------------------------------%
where $\partial_i$ means the partial derivative with respect to the variables $i=r, \theta, z$. The integral in Eq.~(\ref{eq:lagrangian}) should be taken over
the whole elastic medium. The first term of the integrand in Eq.~(\ref{eq:lagrangian}) represents the kinetic energy density, and the other two terms account for the potential energy density with opposite sign.

To the aim, we impose stress-free boundary conditions on the whole surface
of the medium and follow the procedure described
in Ref.~\cite{visscher}.
Expand $u_i$ by a set of basis functions $\phi_\alpha(\bm{r})$:
\begin{equation}%-----------------------------%
 u_i(\bm{r})=\sum_\alpha A_{\alpha i} \phi_\alpha(\bm{r}),
 \label{eq:displacement}
\end{equation}%-------------------------------%
where $\alpha$ labels a basis function. Preferred forms of
$\phi_\alpha(\bm{r})$ depend on the shape of the medium considered,
and those for the present study will be given in Eq.~(\ref{eq:basis}).
Substituting Eq.~(\ref{eq:displacement}) into Eq.~(\ref{eq:lagrangian})
and employing the variation method with respect to $A_{\alpha i}$,
we obtain the generalized eigenvalue equation:
\begin{equation}
 \sum_{\alpha,j} \Gamma_{(\beta i),(\alpha j)} A_{\alpha j}=
 \omega^2 \sum_{\alpha,j} E_{(\beta i),(\alpha j)}A_{\alpha j}.
 \label{eq:general}
\end{equation}
The matrix elements of $E_{(\beta i),(\alpha l)}$
have a simple form written as
\begin{equation}
 E_{(\beta i),(\alpha j)} = \delta_{ij}\frac{1}{V}
 \int_V \phi_{\beta}^{*} \phi_{\alpha} dV.
\label{eq:e_part}
\end{equation}
Expressions of $\Gamma_{(\beta i),(\alpha l)}$
are somewhat complicated, thus summarized
in Appendix \ref{sec:appendix_a}.

\subsection{\label{subsec:level2_b} HOLLOW TUBE WITH MODULATED RADIUS}
Next we restrict our attention to elastic thin hollow tubes with
thickness $2h$ and length $L$, which we consider as a continuum model
of the 1D C$_{60}$ polymers. The tube axis points to the $z$ direction,
and the tube radius is spatially modulated as
\begin{equation}%-----------------------------------------%
 R(z)=R_0 - \frac{a}{2} + \frac{a}{2}
 \cos\left( \frac{2\pi z}{s} \right), \label{eq:radius}
\end{equation}%-------------------------------------------%
where $R_0$ and $R_0-a$ are the maximum and minimum radii, respectively.
Since $R(z)$ is periodic with a period $s$, we can define a unit cell
by a portion of the tube within the region $z\in [0,s]$
as depicted in Fig.~\ref{fig:tube}. For later use,
we introduce a new variable $\xi(z)$ defined by
\begin{equation}
 \xi (z) = \int_{0}^{z} w(z') dz',
 \ \ \ w(z)=\sqrt{1+\left(\partial_z R \right)^2}. \label{eq:xi}
\end{equation}
Here, $\xi (z)$ measures the length of
a geodesic curve \cite{shima_naka} between two points along
the curved membrane represented in Fig.\ref{fig:tube}.
In terms of $\xi (z)$, a unit cell of the tube is a portion
within the region $\xi\in [0,\Lambda]$ where $\Lambda =\xi(s)$.

Once we specify the system's geometry as above,
the volume integral in Eq.~(\ref{eq:e_part}) is written by
\begin{equation}
 \int_V dV = \int_0^{L} dz \int _{R(z)-h}^{R(z)+h} rdr
 \int _{0}^{2\pi} d\theta.  \label{eq:V_out}
\end{equation}
To eliminate the finite length effect, we suppose
$u_i(r,\theta,z)=u_i(r,\theta,z+L)$ with $L=Ns$,
where $N$ is the number of unit cells and $N=800$ in our calculation.
Under the condition, the wavenumber $q$ along the $z$ direction
serves as an additional index (complementary to the label $\alpha$)
that specifies phonon modes; here $q=2\pi p/(N \Lambda)$ with
$-N/2 < p \leq N/2 $ ($p$ is an integer).
The basis function with $q$ and $\alpha$ is defined by
\begin{equation}
 \phi_{q,\alpha} (\bm{r}) = \left( \frac{r}{R_0} \right)^{l} 
 e^{im\theta} e^{i(q-G)\xi},  \label{eq:basis}
\end{equation}
where $l=0, 1, 2,\cdots l_{\mathrm{max}}$, $m=0,\pm 1, \pm 2, \cdots$, and
$G=2\pi n/\Lambda$ $(n=0,\pm 1, \pm 2, \cdots)$.
We used in Eq.~(\ref{eq:basis}) the expression of $e^{iq\xi}$
rather than $e^{iqz}$, since phonons propagate along the $\xi$ axis,
not the $z$ axis. Expression (\ref{eq:basis}) implies that
the label $\alpha$ represents the set of three integers: $(l,m,n)$.

It is noteworthy that $\phi_{q,\alpha}$ with $m$ and
$\phi_{q,\beta}^{*}$ with $m'(\ne m)$ are orthogonal
in the sense of the volume integral (\ref{eq:V_out}).
Therefore, individual phonon dispersion curves in the $q$-$\omega$ plane
are obtained by solving the eigenvalue equation (\ref{eq:general})
with $m$ being fixed.

\subsection{\label{subsec:level2_c} NUMERICAL PARAMETERS}
In the phonon mode analysis, it is essential to determine appropriate values of the Lame coefficients $\lambda, \mu$ and the mass density $\rho$ introduced in Eq.(\ref{eq:lagrangian}). The Lame coefficients are expressed by \cite{landau}
\begin{equation}
 \lambda = \frac{\nu E}{(1+\nu)(1-2\nu)}, \ \ \ 
 \mu = \frac{E}{2(1+\nu)}, \label{eq:elastic_const}
\end{equation}
where $E$ and $\nu$ are the Young modulus and the Poisson ratio of the system, respectively. To our knowledge, there has been no attempt of determining $E$, $\nu$, and $\rho$ of 1D peanut-shaped C$_{60}$ polymers. Meanwhile, pristine C$_{60}$ molecules were suggested to have $E=1.06$TPa, $\nu=0.145$, and $\rho=2.27$g/cm$^3$, whose values are similar to those of graphene sheets \cite{kahn}. We postulate that the three material constants of the 1D C$_{60}$ polymer show the similar values to those noted above, motivated by the fact that adjacent C$_{60}$ molecules in the polymer are connected via the $\sigma$ bonds as same as adjacent carbon atoms within a pristine C$_{60}$ molecule.

Given below is the list of geometric parameters we have used: $R_0=3.5$\AA, $s =7.0$\AA, and $a=1.5$\AA, all of which were referred to the experimental characterization of the 1D C$_{60}$ polymers discussed in Ref.~\cite{takashima1}. The tube thickness is fixed to be $2h=3.2$ \AA \ as derived from the relation $\rho=M_t/v$, where $M_t=1.2\times 10^{-24}$kg is the total mass of 60 carbon atoms and $v= 4\pi \left[ (R_0+h)^3- (R_0-h)^3 \right]/3$ is the volume of a hollow spherical shell with thickness $2h$ that corresponds to the pristine C$_{60}$ atomic structure when it is smeared out.

Before addressing the polymer's dispersion, we have checked the validity of our formulation by applying it to a straight hollow cylinder with no radius modulation. The exact solution of the vibrational eigenmode in the straight cylinder is $e^{i(m\theta + qz)}$ multiplied by $r$-dependent Bessel functions (and their derivatives) \cite{chico}. In our formulation, the $r$-dependent Bessel terms are approximated by a set of polynomial functions, as seen from Eqs.~(\ref{eq:displacement}) and (\ref{eq:basis}). Despite this approximation, the computed phonon frequencies were in agreement with the exact solution to more than 99$\%$ accuracy when we set $l_{\mathrm{max}}=6$. The same value of $l_{\mathrm{max}}$ has been used in the calculations of the modulated cylinders (i.e., the 1D C$_{60}$ polymers), wherein 35 basis functions were employed to solve the eigenvalue equation (\ref{eq:general}); in fact, we set $0 \le l \le 6$ and $-2 \le n \le 2$ for each fixed $m$. It should be noted that for very high frequency range ($\omega>700$ cm$^{-1}$ in our study; see Sec. \ref{sec:level3}), $l_{\mathrm{max}}$ must be larger than our choice, but such the high frequency range is beyond the scope of the present analysis.

%%%%%%%%%%%%%%%%%%%%%%%%%%%%    GRAPHICS   %%%%%%%%%%%%%%%%%%%%%%%%%%%%%%
\begin{figure}[t]
\center
\includegraphics[scale=0.32,clip]{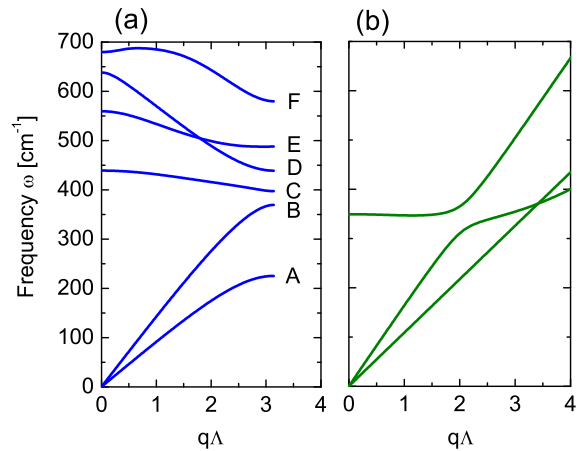}%-----------------------------%
\caption{\label{fig:bands}(Color online)
Phonon dispersion relations of hollow cylinders with periodically varying radius (a) and those with constant radius (b). The angular momentum index $m$ is set to be $m=0$ for both plots. Six capitals (A-F) are assigned to the eigenmodes at $q\Lambda=\pi$ for each dispersion curve in (a).}
\end{figure}
%%%%%%%%%%%%%%%%%%%%%%%%%%%%%%%%%%%%%%%%%%%%%%%%%%%%%%%%%%%%%%%%%%
\section{\label{sec:level3} DESPERSION RELATIONS AND MODE ANALYSIS}
%%%%%%%%%%%%%%%%%%%%%%%%%%%%%%%%%%%%%%%%%%%%%%%%%%%%%%%%%%%%%%%%%%%%
\begin{figure}[tt]
\center
\includegraphics[scale=0.32,clip]{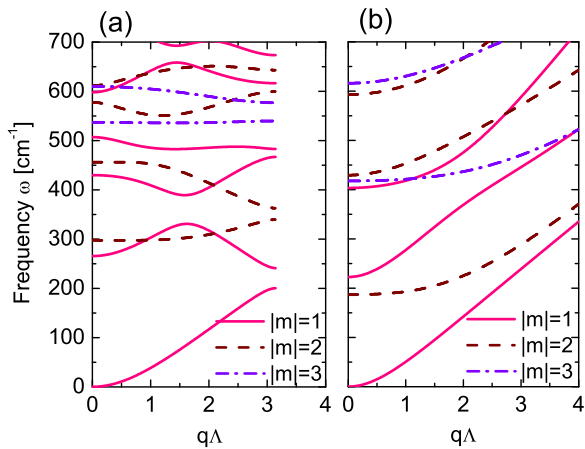}%-----------------------------%
\caption{\label{fig:bands2}(Color online)
Dispersion relations for $\vert m \vert =1,2,3$ of hollow cylinders with: (a) periodically varying radius and (b) constant radius.
}
\end{figure}
%%%%%%%%%%%%%%%%%%%%%%%%%%%%%%%%%%%%%%%%%%%%%%%%%%%%%%
%%%%%%%%%%%%%%%%%%%%%%%%%%%%%%%%%%%%%%%%%%%%%%%%%%%%%%%%%%%%%%%%%%%%
\begin{figure}[tt]
\center
\includegraphics[scale=0.4,clip]{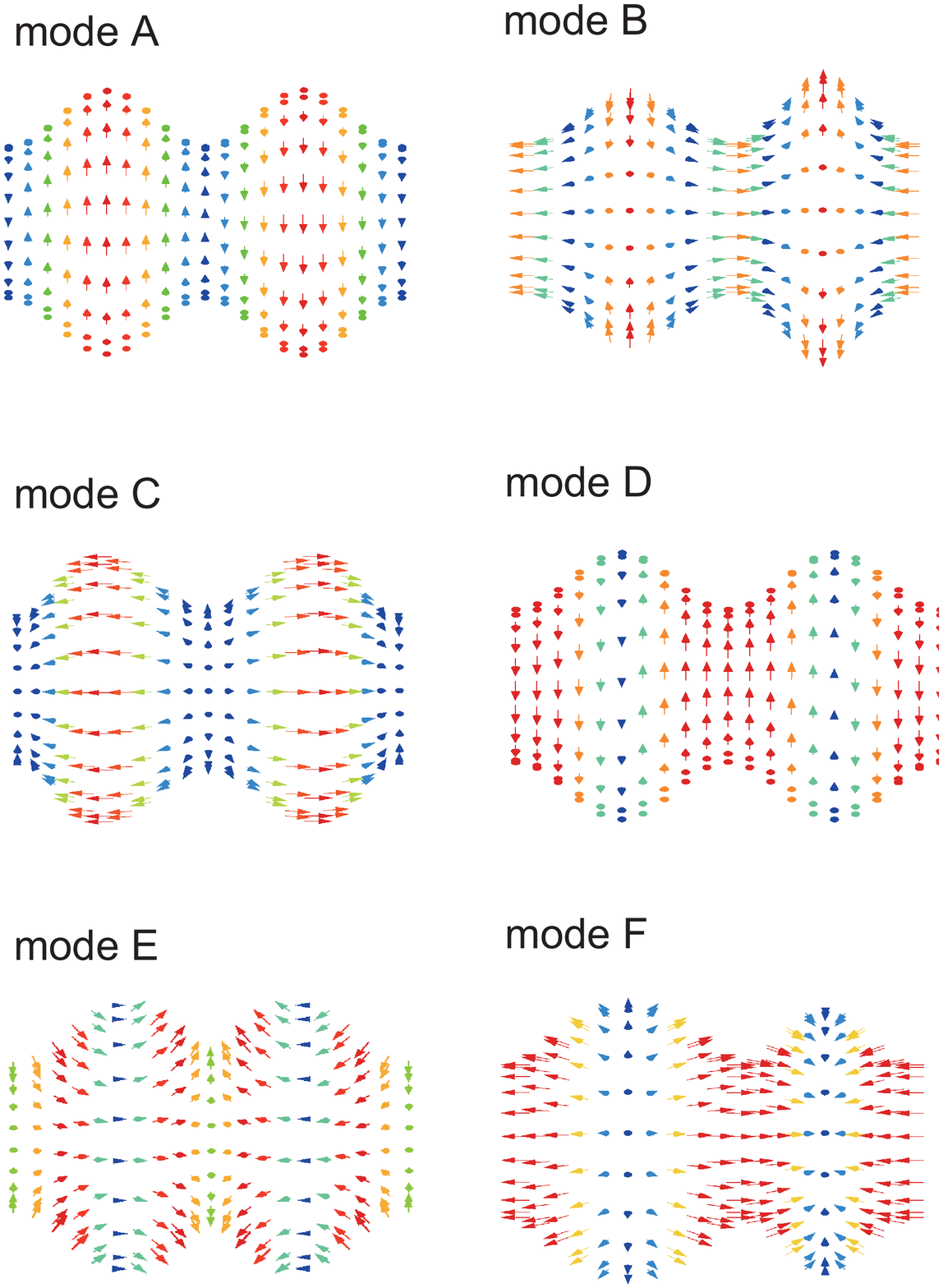}%-----------------------------%
\caption{\label{fig:modes}(Color online)
Displacement distribution of the lowest six modes at $q\Lambda=\pi$ within the region $z \in [s/2,5s/2]$. The magnitude of displacement vectors is signified by the lengths and colors of arrows (red arrows are longer than blue ones in color).}
\end{figure}
%%%%%%%%%%%%%%%%%%%%%%%%%%%%%%%%%%%%%%%%%%%%%%%%%%%%%%%%%%%
%%%%%%%%%%%%%%%%%%%%%%%%%%%%%%%%%%%%%%%%%%%%%%%%%%%%%%%%%%%
Figure \ref{fig:bands}(a) shows the lowest six phonon dispersion curves for $m=0$ in the modulated tube. For comparison, we plot in Fig.~\ref{fig:bands}(b) the lowest three curves for the straight tube with radius $3.5$\AA; the six curves in Fig.~\ref{fig:bands}(a) reduce to the three ones in Fig.~\ref{fig:bands}(b) if we set $a=0$\AA \ [see Eq.~(\ref{eq:radius})]. In the low frequency range ($\omega<200$ cm$^{-1}$), the polymer's dispersion relations are almost identical to those of the straight tube; two almost-linear dispersions are present. Among the two, the bottom curve belongs to a twisting mode ($\omega=c_t q, \ c_t=\sqrt{\mu/\rho}\simeq 14.3$km/s), and the upper to a longitudinal mode ($\omega=c_l q, \ c_l=\sqrt{(\lambda+2\mu)/\rho}\simeq 22.1$km/s). Similar two linear branches were found in carbon nanotubes through continuum \cite{suzuura,chico} and atomistic calculations \cite{gunlycke,zimmermann}, while the magnitude of frequencies deviates from the present results. An important observation in Fig.~\ref{fig:bands}(a) and \ref{fig:bands}(b) is a significant difference in the dispersion profiles at intermediate frequencies ($200$cm$^{-1}<\omega<700$cm$^{-1}$) between the polymer and the straight tube. The periodic radius modulation causes zone-folding at $q\Lambda=\pi$, resulting in additional branches in this frequency range. This implies that the radius modulation effect can be traced by investigating the phononic properties in the branches.

Figures \ref{fig:bands2}(a) and \ref{fig:bands2}(b) show the phonon dispersion relations of the modulated and straight tubes, respectively, for $\vert m \vert = 1,2,3$. Similarly to the case of $m=0$, zone-folding at $q\Lambda=\pi$ and marked difference in the dispersion profiles at the intermediate frequencies are observed. In the following, we specify all the vibrational states in the plots by three indexes: $q$, $m$, and the band index $M=1,2,\cdots$, which labels the $M$th lowest curve for each $m$.

Figure \ref{fig:modes} shows the displacement distribution
in the vibrational modes at $q\Lambda=\pi$ for each six curves
presented in Fig.~\ref{fig:bands}(a). The profiles within the region
$z \in [s/2,5s/2]$ are picked up for all the six modes that we
labeled by A - F in order of increasing the eigenfrequency.
The modes A and D correspond to twisting modes
in which all vectors are parallel to the $\theta$ direction.
The other four are hybridized modes of
longitudinal and radial breathing modes, where vectors point to
directions in-between the $z$ and $r$ axes.

\section{\label{sec:level4} ELECTRON-PHONON INTERACTION}
In view of electron-phonon couplings, the hybridized modes of C and F depicted in Fig.~\ref{fig:modes} give an important consequence. To explain it, we formulate in this section the interaction Hamiltonian between electrons and low-frequency phonons along the deformation potential theory \cite{kittel}. The theory states that low-frequency phonons perturb electronic states through the occurrence of a deformation potential represented by $D\nabla \cdot \bm{\tilde{u}}$. Here, $D$ is the deformation potential constant and $\bm{\tilde{u}}$ denotes the normalized displacement vector. The interaction Hamiltonian is given by
\begin{equation}
 \mathcal{H}_{\mathrm{e-ph}} = \int \Psi^{\dagger}(\theta,z)
 \left[D \nabla \cdot \bm{\tilde{u}} \right]_{r=R(z)}
 \Psi(\theta,z) dS, \label{eq:e-ph_real}
\end{equation}
where $\Psi(\Psi^{\dagger})$ represents the annihilation (creation) operator for electrons. The value of $D$ lies within the range 10-30 eV  for various nanocarbon materials \cite{hwang,leturcq}, although its exact value for the 1D C$_{60}$ polymer has not yet been examined; we thus set $D=20$ eV in the present work. The surface integral in Eq.~(\ref{eq:e-ph_real}) is written as
\begin{equation}
 \int dS= \int_{0}^{2\pi} d\theta \int_{0}^{Ns} dz R(z) w(z).
\end{equation}

We now derive an alternative form of $\mathcal{H}_{\mathrm{e-ph}}$
[i.e., Eq.~(\ref{eq:e-ph_reciprocal})] by expanding
the field operators $\bm{\tilde{u}}$ and $\Psi$ introduced
in Eq.~(\ref{eq:e-ph_real}). First,
the $i$th component of the quantized phonon field
$\bm{\tilde{u}}$ is expressed by
\begin{equation}
 \tilde{u_i}=\sum_J 
 \left(b_J u_{J,i} + b_{J}^{\dagger} u_{J,i}^{*} \right),
 \label{eq:phonon_field}
\end{equation}
where $b_J (b_{J}^{\dagger})$ is the phonon annihilation (creation) operator
and $J$ denotes the mode index of phonons, i.e., $\omega_J = \omega_{m,q,M}$.
The $u_{J,i}$ in Eq.~(\ref{eq:phonon_field}) is written as
\begin{equation}
 u_{J,i}=c_J \sum_{\alpha} A_{\alpha i}^{(J)}
       \phi_{\alpha} (\bm{r}).
 \label{eq:normalization}
\end{equation}
Following the quantization procedure~\cite{nishiguchi2},
$c_J$ should be determined by carrying out the integral:
\begin{equation}
 \int \bm{u}_{J'}^* \cdot \bm{u}_{J} dV = \frac{\hbar}{2\rho\omega_J}
 \delta_{J,J'}. \label{eq:quantization}
\end{equation}
Consequently, we obtain the normalization constants:
\begin{equation}
c_J=\frac{1}{\sqrt{\bm{A}^{(J)\dagger}
           \cdot\bm{E}\cdot\bm{A}^{(J)}}}
           \sqrt{\frac{\hbar}{2\rho V \omega_J}},
\end{equation}
where $\bm{A}^{(J)}$ is the eigenvector corresponding to $\omega_J$,
$\bm{E}$ is the matrix defined in Eq.~(\ref{eq:general}),
$V$ is the volume of the system.

\begin{figure}[tt]
\center
\includegraphics[scale=0.34,clip]{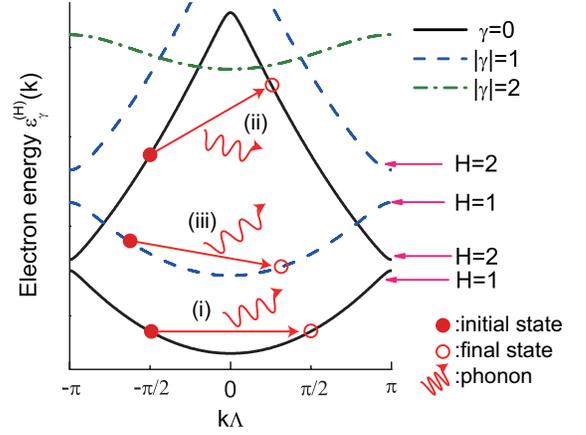}%-----------------------------%
\caption{\label{fig:scattering}(Color online)
Electron band structure and phonon emission process for several bands.
The intra-band electron scatterings of (i), (ii), and (iii) within the electronic band labeled by ($\vert \gamma \vert$, $H$)=(0,1), (0,2), and (1,1) are depicted, respectively.
}
\end{figure}

Next, the field operator for the electron is written as
\begin{equation}
 \Psi(\theta,z)=\sum_{\gamma,k,H,\sigma} a_{\gamma,k,\sigma}^{(H)}
 \psi_{\gamma,k}^{(H)}(\theta,z), \label{eq:electron_field}
\end{equation}
where $a_{\gamma,k,\sigma}^{(H)}$ is the annihilation operator
of the electron with spin $\sigma$, $\gamma$ is the angular quantum
number, $k$ is the wavenumber, and $H$ is the band index.
Explicit forms of the electron's eigenfunctions
$\psi_{\gamma,k}^{(H)}(\theta,z)$ are derived in Appendix \ref{sec:appendix_b}.
The electronic band structure $\epsilon_{\gamma}^{(H)}$ as a function of $k$
is obtained by numerical diagonalizations of the Schr\"{o}dinger
equation (\ref{eq:electron}). The lowest few electronic dispersion curves
$\epsilon_{\gamma}^{(H)}(k)$ for several $\gamma$ and $H$
are drawn in Fig.~\ref{fig:scattering}. Zone-folding at $k\Lambda=\pm \pi$
is due to the curvature-induced electrostatic potential with the same
periodicity of the radius modulation \cite{shima1,fujita}.

%%%%%%%%%%%%%%%%%%%%%%%%%%%%%%%%%%%%%%%%%%%%%%%%%%%%%%%%%%%%%%%%%%%%
%%%%%%%%%%%%%%%%%%%%%%%%%%%%%%%%%%%%%%%%%%%%%%%%%%%%%%%%%%%
Finally, we arrive at the interaction Hamiltonian
in the reciprocal-space representation by substituting
Eqs.~(\ref{eq:phonon_field}) and (\ref{eq:electron_field})
into Eq.~(\ref{eq:e-ph_real}):
\begin{equation}
 \mathcal{H}_{e-ph} = \sum_{\sigma} \sum_{P_i,P_f}\sum_{J}
 g_{J}^{P_i,P_f}
 a_{P_f,\sigma}^{\dagger}a_{P_i,\sigma} (b_{J}+b_{-J}^{\dagger}),
 \label{eq:e-ph_reciprocal}
\end{equation}
where the labels $J=(m,q,M)$ and $P_j=(\gamma_{j},k_j,H_j)$ indicate the phonon and electron modes, respectively. The quantity $g_{J}^{P_i,P_f}$ describes the strength of the electron-phonon interaction through which electrons are scattered from initial states ($j=i$) to final ones ($j=f$); several scattering processes are illustrated in Fig.~\ref{fig:scattering}. We can prove that
\begin{eqnarray}
 & &g_{J}^{P_i,P_f}
 =D c_{J} \sum_{\alpha}
 \Bigg[ A_{\alpha r}^{(J)} \left(\frac{l+1}{R_0}\right)
  K_{l-1,G,0}^{P_i,P_f,m,q}   \nonumber\\
 &+& A_{\alpha z}^{(J)} \left(\frac{im}{R_0}\right)
                    K_{l-1,G,0}^{P_i,P_f,m,q}
  + A_{\alpha z}^{(J)} i (q-G) K_{l,G,1}^{P_i,P_f,m,q}
 \Bigg], \nonumber\\
 \label{eq:e-ph_coupling}
\end{eqnarray}
where
\begin{eqnarray}
  K_{l,G,p}^{P_i,P_f,m,q} &=& \sum_{\tilde{l}}
 \delta_{\gamma_{i}+m,\gamma_{f}}
 \delta_{k_i+q-k_f,2\pi \tilde{l}/\Lambda} \nonumber\\
 &\times& \frac{1}{\Lambda} \sum_{G',G''} C_{\gamma_{f},k_f-G'}^{(H_f)*}
              C_{\gamma_{i},k_i-G''}^{(H_i)} \nonumber\\
  &\times& \int_{0}^{s} dz w^{1+p} 
 \left(\frac{R}{R_0}\right)^{l_p}
 e^{i(2\pi \tilde{l}/\Lambda-G+G'-G'')\xi(z)}, \nonumber\\
 \label{eq:e-ph_K}
\end{eqnarray}
and $C_{\gamma,k-G}^{(H)}$ is a Fourier expansion coefficient of
$\psi_{\gamma,k}^{(H)}$ as given by Eq.~(\ref{eq:electron_wave}).
In Eq.~(\ref{eq:e-ph_K}),
the integer $\tilde{l}$ is introduced to let $k_i$ and $k_f$ be in the
first Brillouin zone, i.e., $k_j\Lambda \in [-\pi,\pi]$. 

%%%%%%%%%%%%%%%%%%%%%%%%%%%%%%%%%%%%%%%%%%%%%%%%%%%%%%%%%%%%%%%%%%%%
\begin{figure}[tt]
\center
\includegraphics[scale=0.34,clip]{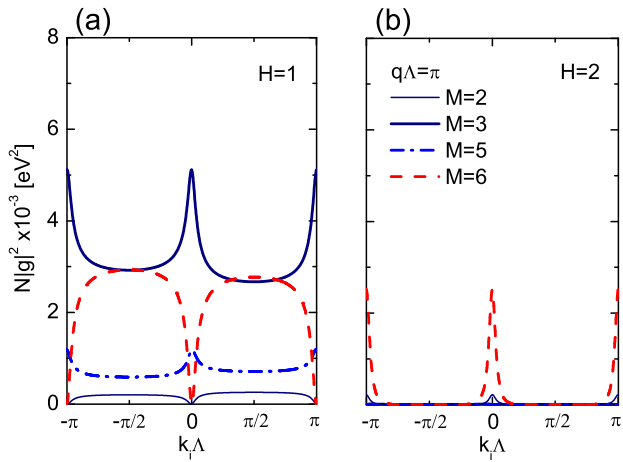}%-----------------------------%
\caption{\label{fig:coupling1}(Color online)
Coupling strength $\vert g \vert^2$ in the case of $\gamma=0$. The dependences on the electron's initial wavenumber $k_i$ are displayed, while the phonon mode parameters $m=0$ and $q\Lambda=\pi$ are fixed.}
\end{figure}
%%%%%%%%%%%%%%%%%%%%%%%%%%%%%%%%%%%%%%%%%%%%%%%%%%%%%%%%%%%
\section{\label{sec:level5} COUPLING STRENGTH}
\begin{figure}[tt]
\center
\includegraphics[scale=0.34,clip]{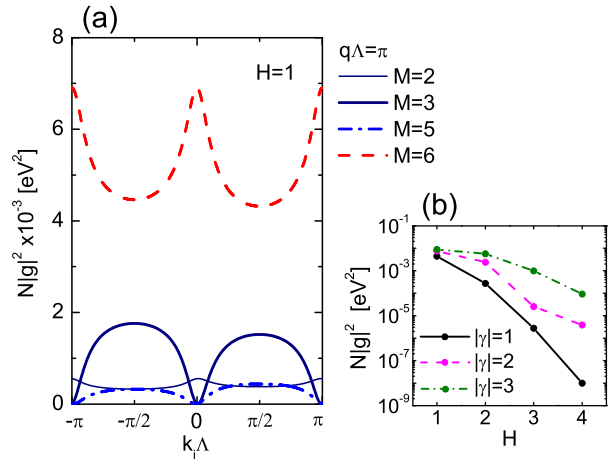}%-----------------------------%
\caption{\label{fig:coupling2}(Color online)
(a) The $k_i$-dependence of $\vert g \vert^2$ for $\vert \gamma \vert = 1$. (b) Rapid decrease in $\vert g \vert^2$ of $M=6$ at $k_i\Lambda=\pi/2$ with the electron band index $H$.
}
\end{figure}

The coupling strength $\vert g_{J}^{P_i,P_f} \vert^2$ for various conditions of $J (=m,q,M)$ and $P_j (= \gamma_j,k_j,H_j)$ with $j=i~\mathrm{or}~f$ has been systematically calculated, only a part of which are presented in Figs.~\ref{fig:coupling1} and \ref{fig:coupling2}. In all the figures, we set $m\equiv 0$, $q\Lambda\equiv\pi$, $\gamma_i=\gamma_f(\equiv \gamma)$, and $H_i=H_f(\equiv H)$, since the intra-band electron scatterings that satisfy these conditions (see Fig.~\ref{fig:scattering}) were found to yield significantly large values of $\vert g_{J}^{P_i,P_f} \vert^2$. Hereafter, the suffixes in $g_{J}^{P_i,P_f}$ are omitted for simple notation.

First, we look into $\vert g \vert^2$ for $\gamma = 0$ and $H=1$, in which only
the process (i) depicted in Fig.~\ref{fig:scattering} is relevant.
Figure~\ref{fig:coupling1}(a) shows $\vert g \vert^2$
as a function of the initial wavenumber $k_i$ for four different $M$s,
i.e., $P_i=(0,k_i,1)$, $P_f=(0,k_i+\pi/\Lambda,1)$, and $J=(0,\pi/\Lambda,M)$.
The contributions from $M=1,4$ to $\vert g \vert^2$
vanish at every $k_i$, because these twisting modes give $\nabla\cdot \bm{\tilde{u}}=0$.
For almost all $k_i$, $\vert g \vert^2$ with $M=3$ takes the largest value,
while $\vert g \vert^2$ with $M=6$ competes within limited ranges around
$k_i\Lambda \simeq \pm \pi/2$.
The primary contributions to $\vert g \vert^2$ from the phonon modes
of $M=3,6$ can be explained by their longitudinal nature.
As shown in Fig.~\ref{fig:modes},
the displacement vectors in the two modes (C and F) are mostly tangential
to the modulated cylindrical surface, thus
yielding large values of $\nabla \cdot \bm{\tilde{u}}$. 
This results in the enhancement of the electron-phonon coupling as follows
from the expression of $\mathcal{H}_{\mathrm{e-ph}}$ given
in Eq.~(\ref{eq:e-ph_reciprocal}).

The profile of $\vert g \vert^2$ drastically changes when electrons stay at upper bands of $H\ge 2$. Figure~\ref{fig:coupling1}(b) shows the contribution to $\vert g \vert^2$ from the process (ii) in Fig.~\ref{fig:scattering},
i.e., $H=2$ and $\gamma =0$. In this case, $\vert g \vert^2$ is strongly depressed except for a few peaks at $k_i\Lambda=0, \pm \pi$. If we increase $H$ further, the peaks diminish rapidly and $\vert g \vert^2$ disappears for all $k_i$. This disappearance is in contrast with $\vert g \vert^2$ for $H=1$, where every $k_i$ give finite values of $\vert g \vert^2$. We thus conclude that, in the modulated cylinders, significant values of $\vert g \vert^2$ can be obtained only when the coupled electrons belong to the lowest band of $H=1$.
%%%%%%%%%%%%%%%%%%%%%%%%%%%%%%%%%%%%%%%%%%%%%%%%%%%%%%%%%%%%%%%%%%%%
%%%%%%%%%%%%%%%%%%%%%%%%%%%%%%%%%%%%%%%%%%%%%%%%%%%%%%%%%%%

Next, we pay attention to the case of $\gamma \ne 0$. Figure~\ref{fig:coupling2}(a) shows $\vert g \vert^2$s for $\vert \gamma \vert =1$ and $H=1$ [the process (iii) in Fig.~\ref{fig:scattering}], in which $\vert g \vert^2$ with $M=6$ is the largest for all $k_i$ and that with $M=3$ shows a modest value. We have confirmed that as $H$ increases, every curves of $\vert g \vert^2$ go downward, which result in profiles of $\vert g \vert^2$ analogous to Fig.~\ref{fig:coupling1}(b). Similar depression of $\vert g \vert^2$ with increasing $H$ have been observed for $\vert \gamma \vert \ge 2$, as shown in Fig.~\ref{fig:coupling2}(b). We can say that for arbitrary $\gamma$, $\vert g \vert^2$ for $M=3$ or $6$ has significant value only when both two coupled electrons reside at a small $H$ branch.

\section{\label{sec:level6} PHYSICAL CONSEQUENCES}
%%%%%%%%%%%%%%%%%%%%%%%%%%%%%%%%%%%%%%%%%%%%%%%%%%%%%%%%%%%%%%%%%%%%
\begin{figure}[tt]
\center
\includegraphics[scale=0.34,clip]{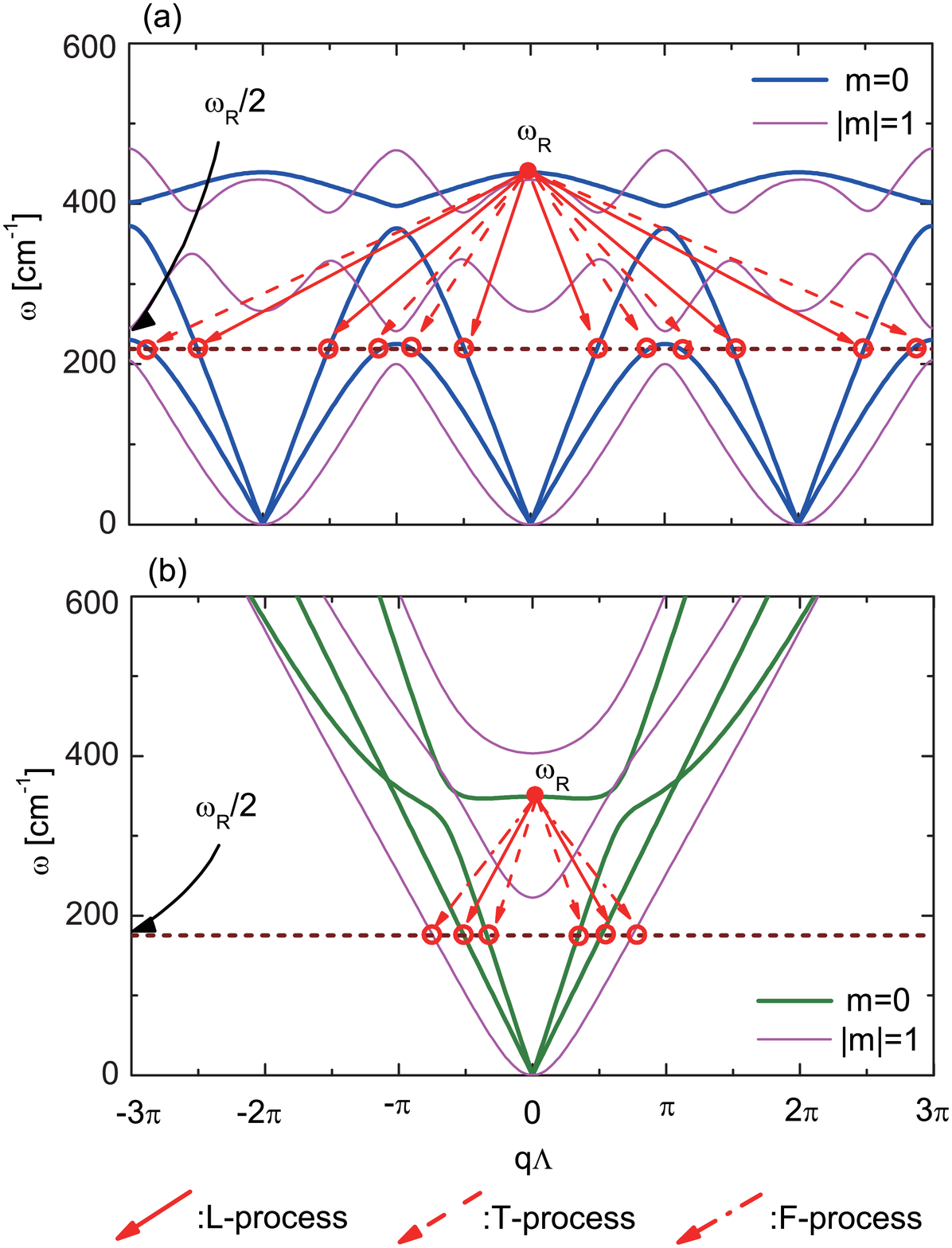}
\caption{\label{fig:decay}(Color online)
Allowed relaxation process of RBMs for (a) the 1D C$_{60}$ polymer and (b) a straight tube. The L-, T-, and F-processes suggest the RBM's relaxation in which the post-decay modes are the longitudinal, twisting, and flexural modes, respectively.
}
\end{figure}
%%%%%%%%%%%%%%%%%%%%%%%%%%%%%%%%%%%%%%%%%%%%%%%%%%%%%%%%%%%
\subsection{DOMINANT MODE COUPLING}
Our findings of $\vert g \vert^2$ in Sec.\ref{sec:level5} give an important suggestion for addressing the property of the possible Peierls transition in the 1D C$_{60}$ polymers. As is well known, the Peierls transition temperature $T_c$ in general 1D systems is dependent on the coupling strength $\vert g \vert^2$; the larger value $\vert g \vert^2$ arises, the higher $T_c$ we observe \cite{gruner}. It follows from Sec.~\ref{sec:level5} that $\vert g \vert^2$ becomes significant only when the electrons lie at a small $H$ branch and a specific phonon mode of $M=3$ or $6$ is involved. All the rest combination of electronic and phononic modes give slight contribution to the Peierls transition in the 1D C${_{60}}$ polymers. In this context, we have succeeded to determine what specific modes (electron and phonon) in the 1D C${_{60}}$ polymers should be responsible for the Peierls transition endowed with a feasible value of $T_c$. Quantitative evaluation of $T_c$ for the 1D C$_{60}$ polymers is possible via the standard Green's function technique \cite{abri} incorporated with our formulation of $\vert g \vert^2$ given in Eqs.~(\ref{eq:e-ph_coupling}) and (\ref{eq:e-ph_K}). The calculated results of $T_c$ as well as the comparison with experiments will be discussed elsewhere.

%%%%%%%%%%%%%%%%%%%%%%%%%%%%%%%%%%%%%%%%%%%%%%%%%%%%%%%%%%%
We emphasize that the present formulation of $\vert g \vert^2$ is broadly applicable to the study of the radius modulation effects on the physical properties of 1D C$_{60}$ polymers. The application should include the electrical resistivity analyses. In general, the low-temperature resistivity of nanomaterials is dependent on temperature and the Fermi energy, and the dependences are governed largely by the magnitude of $\vert g \vert^2$. Hence, our formulation of $\vert g \vert^2$ will be helpful to estimate the low-temperature resistivity of the 1D C$_{60}$ polymers that remains untouched to date. In addition, artificial control of the Fermi level by chemical doping or bias voltage application may enable to obtain a desired magnitude of $\vert g \vert^2$, leading to an efficient means of developing next-generation nanodevices.

%%%%%%%%%%%%%%%%%%%%%%%%%%%%%%%%%%%%%%%%%%%%%%%%%%%%%%%%%%%
Besides the modulated cylinder problem, an extension of the present theory to three dimensional counterparts (e.g., the Mackay crystal \cite{mackay} and the zeolite-templated carbon \cite{takai}) is quite interesting. The Mackay crystal is a three-dimensional carbon network; three infinite tubes intersect smoothly in the unit cell, resulting in a triply periodic minimal surface. Due to the intriguing geometry and functionality \cite{park}, it has attracted interest since the first hypothesis. We speculate that in the Mackay crystal, only specific phonon modes cause large deformation potential similarly to the present system, since each constituent tube in the three dimensional minimal surface can be regarded as a radius-modulated tube along the axial direction. Electronic structures of the periodic minimal surface have been already clarified \cite{aoki,koshino,fujita2}, and thus employing the results to obtain the electron-phonon interaction Hamiltonian will yield a conclusion of this issue.

%%%%%%%%%%%%%%%%%%%%%%%%%%%%%%%%%%%%%%%%%%%%%%%%%%%%%%%%%%%
\subsection{RADIAL BREATHING MODE RELAXATION}
Another possible consequence of the periodic radius modulation in 1D C$_{60}$ polymers is a considerable decrease in the relaxation time of radial breathing modes (RBMs) compared with that in carbon nanotubes \cite{rao3}. We have numerically found that the phononic mode labeled by ($m,q,M$)=($0,0,3$) is a RBM, which is active on the resonant Raman scattering. The lifetime of a RBM is mainly determined by phonon-phonon scatterings which lead to a phonon decay \cite{rao3}. When the number of the allowed decay process is large, the corresponding phonon mode has a short life. This indeed holds true for the RBM of the 1D C$_{60}$ polymers, as demonstrated below.

Figures~\ref{fig:decay}(a) and \ref{fig:decay}(b) illustrate the RBM decay processes in the 1D C$_{60}$ polymer and the straight tube, respectively. Allowed decay processes are represented by slanted arrows that extend from the RBM (indicated by $\omega_R$) to low-lying acoustic modes having the eigenfrequency $=\omega_R/2$: The post-decay states locate at the intersections of the horizontal dotted line $(=\omega_R/2)$ and dispersion curves. Note that the decay process is constrained by the conservation laws of energy, wavenumber, and angular momentum. Therefore, a RBM having the eigenfrequency $\omega_R$ decays into a pair of acoustic phonons with $\omega_{m,q,M}$ and $\omega_{-m,-q+G,M}$ that satisfy $\omega_{R}=\omega_{m,q,M}+\omega_{-m,-q+G,M}$; here, $G=0$ for normal processes and $G\ne 0$ for umklapp processes. It is found that in the straight tube ($\omega_R=350$cm$^{-1}$), there are four decay processes: L-, T-, and doubly-degenerate F-processes (i.e., $\omega_{1,q,1}+\omega_{-1,-q,1}$ and $\omega_{1,-q,1}+\omega_{-1,q,1}$), where L, T, and F mean that the post-decay modes are the longitudinal, twisting, and flexural modes, respectively. All the processes belong to the normal process (i.e., $G=0$). On the other hand, the RBM of the polymer ($\omega_R=440$cm$^{-1}$) can decay through both the normal and umklapp processes ($G\ne 0$) involving L- and T-processes. Due to many choices of $G$ available, the number of decay process in the polymer is much larger than that in the straight tube, although the F-process in the polymer is forbidden as the dispersion curves with $\vert m \vert=1$ has the frequency gap around $\omega_R/2$. In other words, the zone-folding in the phonon dispersion results in a considerable increase in the number of the decay process in the polymer case, which leads to a broadening of the RBM peak linewidth. It is known that at room temperature, the linewidth of the RBM peak in carbon nanotubes is $\sim 1$cm$^{-1}$ independently of the tube radius \cite{rao3}. Hence, that in the 1D C$_{60}$ polymer is expected to be larger than 1cm$^{-1}$ as a result of the periodic radius modulation. The theoretical estimation of the linewidth will be carried out in our future work.

\section{CONCLUSION}
\label{sec:level7}
We have investigated the low-frequency phonon states and their coupling
with electrons in the 1D peanut-shaped C$_{60}$ polymers along the continuum approximation. It has been found that the phonon dispersion relations exhibit
the zone-folding effect due to the periodic radial modulation.
The electron-phonon interaction was formulated in line with the deformation
potential theory, and the obtained formula for the coupling strength
allows us to identify which phonon modes should be responsible for
the Peierls transition observed in the prior experiment.
The present formulation for the coupling strength has a wide applicability for estimating physical quantities of the systems, thus playing a crucial role in predicting untouched behaviors governed by the electron-phonon interaction.

%%%%%%%%%%%%%%%%%%%%%%%%%%%%%%%%%%%%%%%%%%%%%%%%%%%%%%%%%%%%%%%%%%%%%%%%%%%%%%%
\begin{acknowledgments}
We would like to thank K. Yakubo, N. Nishiguchi, S. Mizuno and Y. Tanaka
for helpful suggestions toward formulation of the elasticity theory.
We also appreciate informative comments given by H. Suzuura and H. Yoshioka.
This study was supported by a Grant-in-Aid for Scientific
Research from the MEXT, Japan. HS acknowledgments the supports from Hokkaido
Gas Co., Ltd. and The Sumitomo Foundation.
A part of numerical simulations were carried out
using the facilities of the Supercomputer Center, ISSP, University of Tokyo.
\end{acknowledgments}
%%%%%%%%%%%%%%%%%%%%%%%%%%%%%%%%%%%%%%%%%%%%%%%%%%%%%%%%%%%%%%%%%%%%%%%%%%%%%%%

\appendix

\section{EXPRESSIONS OF $\Gamma_{(\beta i)(\alpha l)}$}
\label{sec:appendix_a}
The matrix elements $\Gamma_{(\beta i),(\alpha l)}$ introduced in
Eq.~(\ref{eq:general}) are represented as follows.
\begin{widetext}
%%%%%%%%%%%%% ----- r part ------- %%%%%%%%%%%%%%%%%%%%%%%%%%%%%%%
For $\Gamma_{(\beta i),(\alpha l)}$ with $i \equiv r$, we have
\begin{eqnarray}
 \Gamma_{(\beta r),(\alpha r)} &=& \frac{\lambda}{\rho V}
 \int \Phi_{\beta,+}^{*}  \Phi_{\alpha,+} dV
 + \frac{\mu}{\rho V}
 \int \left(
         2 \partial_r \phi_{\beta}^{*} \partial_r \phi_{\alpha}
  +\frac{1}{r^2}\partial_\theta \phi_{\beta}^{*}
                \partial_\theta \phi_{\alpha}
          +\partial_z \phi_{\beta}^{*} \partial_z \phi_{\alpha}
  +\frac{2}{r^2} \phi_{\beta}^{*} \phi_{\alpha}
     \right)dV, \nonumber\\  %========================================%
 \Gamma_{(\beta r),(\alpha \theta)} &=& \frac{\lambda}{\rho V}
  \int \Phi_{\beta,+}^{*} \frac{1}{r} \partial_\theta \phi_{\alpha} dV
 + \frac{\mu}{\rho V}
 \int \left(
         \frac{1}{r} \partial_\theta \phi_{\beta}^{*}
         \Phi_{\alpha,-}
  +\frac{2}{r^2}\phi_{\beta}^{*} \partial_\theta \phi_{\alpha}
     \right) dV, \nonumber\\   %========================================%
 \Gamma_{(\beta r),(\alpha z)} &=& \frac{\lambda}{\rho V}
  \int  \Phi_{\beta,+}^{*} 
                 \partial_z \phi_{\alpha} dV
 + \frac{\mu}{\rho V}
 \int \partial_z \phi_{\beta}^{*} \partial_r \phi_{\alpha} dV.
\label{eq:r_part}
\end{eqnarray}
%%%%%%%%%%%%% ----- theta part ------- %%%%%%%%%%%%%%%%%%%%%%%%%%%%%%%
For $\Gamma_{(\beta i),(\alpha l)}$ with $i \equiv \theta$,
\begin{eqnarray}
 \Gamma_{(\beta \theta),(\alpha r)} &=& \frac{\lambda}{\rho V}
 \int \frac{1}{r} \partial_\theta \phi_{\beta}^{*}
         \Phi_{\alpha,+} dV
 + \frac{\mu}{\rho V}
 \int \left(
         \frac{2}{r^2} \partial_\theta \phi_{\beta}^{*} \phi_{\alpha}
  + \Phi_{\beta,-}^{*}
     \frac{1}{r} \partial_\theta \phi_{\alpha}
     \right) dV, \nonumber\\  %=========================================%
 \Gamma_{(\beta \theta),(\alpha \theta)} &=& \frac{\lambda}{\rho V}
  \int \frac{1}{r^2}\partial_\theta \phi_{\beta}^{*}
           \partial_\theta \phi_{\alpha} dV
 + \frac{\mu}{\rho V}
 \int \left(
          \frac{2}{r^2}\partial_\theta \phi_{\beta}^{*}
           \partial_\theta \phi_{\alpha}
 + \Phi_{\beta,-}^{*}  \Phi_{\alpha,-}
      +\partial_z \phi_{\beta}^{*} \partial_z \phi_{\alpha}
     \right) dV, \nonumber\\  %=========================================%
 \Gamma_{(\beta \theta),(\alpha z)} &=& \frac{\lambda}{\rho V}
  \int \frac{1}{r} \partial_\theta \phi_{\beta}^{*}
                   \partial_z \phi_{\alpha} dV
 + \frac{\mu}{\rho V}
 \int \frac{1}{r}\partial_z \phi_{\beta}^{*} \partial_\theta \phi_{\alpha} dV.
\label{eq:theta_part}
\end{eqnarray}
%%%%%%%%%%%%% ----- z part ------- %%%%%%%%%%%%%%%%%%%%%%%%%%%%%%%
For $\Gamma_{(\beta i),(\alpha l)}$ with $i \equiv z$,
\begin{eqnarray}
 \Gamma_{(\beta z),(\alpha r)} &=& \frac{\lambda}{\rho V}
 \int  \partial_z \phi_{\beta}^{*} \Phi_{\alpha,+} dV
 + \frac{\mu}{\rho V}
 \int \left(
           \partial_r \phi_{\beta}^{*} \partial_z \phi_{\alpha}
     \right)dV, \nonumber\\
 \Gamma_{(\beta z),(\alpha \theta)} &=& \frac{\lambda}{\rho V}
  \int \frac{1}{r}\partial_z \phi_{\beta}^{*} \partial_\theta \phi_{\alpha} dV
 + \frac{\mu}{\rho V}
 \int \frac{1}{r} \partial_\theta \phi_{\beta}^{*}
                  \partial_z \phi_{\alpha} dV, \nonumber\\
 \Gamma_{(\beta z),(\alpha z)} &=& \frac{\lambda}{\rho V}
  \int \partial_z \phi_{\beta}^{*} \partial_z \phi_{\alpha} dV
 + \frac{\mu}{\rho V}
 \int \left(
           \partial_r \phi_{\beta}^{*} \partial_r \phi_{\alpha}
  +\frac{1}{r^2} \partial_\theta \phi_{\beta}^{*} \partial_\theta \phi_{\alpha}
         +2\partial_z \phi_{\beta}^{*} \partial_z \phi_{\alpha}
     \right) dV.
\label{eq:z_part}
\end{eqnarray}
\end{widetext}
Here, we have used the following expression
\begin{eqnarray}
 \Phi_{p,\pm}=\partial_r \phi_{p} \pm \frac{\phi_{p}}{r}. \ \ [p=\alpha,\beta]
\end{eqnarray}
The volume integral and the basis function $\phi_{\alpha(\beta)}$
are defined in Eqs.~(\ref{eq:V_out}) and (\ref{eq:basis})
in Sec.~\ref{subsec:level2_b}, respectively.

\section{ELECTRON EIGENSTATES}
\label{sec:appendix_b}
In this Appendix, we derive electronic states in the cylindrical tube with varying radius. We have assumed that the tube thickness is small enough to suppress electron's excitation in the normal direction. The same assumption works well for treating the interaction between electrons in both 1D C$_{60}$ polymers \cite{shima1} and two-dimensional curved thin films \cite{ono1}. Due to the rotational symmetry of the tube, the wavefunction becomes $\psi (\theta, z) = \left( e^{i\gamma \theta}/\sqrt{2\pi} \right) \chi_\gamma (z)$ with the electronic angular quantum number $\gamma$. The Schr\"{o}dinger equation for $\chi_\gamma (z)$ is written as \cite{shima1,fujita}
\begin{equation}
 -\frac{\hbar^2}{2m^*}
 \left[\frac{1}{Rw}\partial_z \left(\frac{R}{w}\partial_z \right)
      +U_\gamma (z) \right] \chi_\gamma (z)
      =\epsilon_\gamma \chi_\gamma (z). \label{eq:electron}
\end{equation}
Here, $m^*$ is the effective mass and $U_\gamma (z)$ is an effective potential expressed by
\begin{equation}
 U_\gamma (z) = - \frac{\gamma^2}{R^2}
        + \left( \frac{w^2 - R \partial_{z}^{2} R}{2Rw^3} \right)^2
     +\frac{\partial_{z}^{2} R}{Rw^4}.
\end{equation}
Since $R(z)$ and $w(z)$ are periodic as shown by Eqs.~(\ref{eq:radius}) and (\ref{eq:xi}), $U_\gamma (z)$ is also periodic, i.e., $U_\gamma (z)=U_\gamma (z+s)$. Thus, it follows from Bloch's theorem that
\begin{equation}
 \psi_{\gamma,k}^{(H)}(\theta,z)=\frac{e^{i\gamma\theta}}{\sqrt{2\pi}} \left[
 \frac{1}{\sqrt{N\Lambda R}}\sum_{G_e} C_{\gamma,k-G_e}^{(H)} e^{i(k-G_e)\xi(z)} \right], \label{eq:electron_wave}
\end{equation}
where $G_e=2\pi n_e/\Lambda \ (n_e=0,\pm 1, \pm 2, \cdots)$.

\nocite{*}


\end{document}